\documentclass[pre, amsmath, amssymb, showpacs,
superscriptaddress, twocolumn]{revtex4}
\usepackage{bm}
\usepackage[mathcal]{eucal}
\usepackage{bbm}
\usepackage{nicefrac}
\usepackage{graphicx}
\usepackage{color}

\renewcommand{\vec}[1]{\bm{#1}}
\newcommand{\uvec}[1]{\hat{\vec{#1}}}

\newcommand{\avr}[1]{\left\langle#1\right\rangle}

\newcommand{\Lv}{\mathcal{L}}

\begin{document}

\title{Partitioning of energy in highly polydisperse granular gases}
\author{H. Uecker}
\affiliation{Institute of Theoretical Physics, University of G\"ottingen,
Friedrich-Hund-Platz 1, 37077 G\"ottingen, Germany}
\affiliation{Mathematics and Biosciences Group, Faculty of
  Mathematics, University of Vienna, Nordbergstrasse 15, A-1090 Vienna, Austria}
\author{W. T. Kranz}
\affiliation{Institute of Theoretical Physics, University of G\"ottingen,
Friedrich-Hund-Platz 1, 37077 G\"ottingen, Germany}
\affiliation{Max Planck Institute for Dynamics and Self Organization, 
Bunsenstr. 10, 37073 G\"ottingen, Germany}
\author{T. Aspelmeier}
\affiliation{Institute of Theoretical Physics, University of G\"ottingen,
Friedrich-Hund-Platz 1, 37077 G\"ottingen, Germany}
\affiliation{Max Planck Institute for Dynamics and Self Organization, 
Bunsenstr. 10, 37073 G\"ottingen, Germany}
\author{A. Zippelius}
\affiliation{Institute of Theoretical Physics, University of G\"ottingen,
Friedrich-Hund-Platz 1, 37077 G\"ottingen, Germany}
\affiliation{Max Planck Institute for Dynamics and Self Organization, 
Bunsenstr. 10, 37073 G\"ottingen, Germany}


\date{\today}

\begin{abstract}
  A highly polydisperse granular gas is modeled by a continuous
  distribution of particle sizes, $a$, giving rise to a corresponding
  continuous temperature profile, $T(a)$, which we compute
  approximately, generalizing previous results for binary or
  multicomponent mixtures. If the system is driven, it evolves towards
  a stationary temperature profile, which is discussed for several
  driving mechanisms in dependence on the variance of the size
  distribution. For a uniform distribution of sizes, the stationary
  temperature profile is nonuniform with either hot small particles
  (constant force driving) or hot large particles (constant velocity
  or constant energy driving). Polydispersity always gives rise to
  non-Gaussian velocity distributions. Depending on the driving
  mechanism the tails can be either overpopulated or underpopulated as
  compared to the molecular gas. The deviations are mainly due to
  small particles. In the case of free cooling the decay rate depends
  continuously on particle size, while all partial temperatures decay
  according to Haff's law. The analytical results are supported by
  event driven simulations for a large, but discrete number of
  species.
\end{abstract}

\maketitle

\section{Introduction}
\label{sec:intro}

Granular media are an important and popular subject of current
research which is owed partly to the striking phenomena they reveal
and partly to their ubiquity in nature and in industry which makes a
good understanding of their properties indispensable
\cite{shinbrot+muzzio00,aranson+tsimring06,goldhirsch03}.  Of special
interest are mixtures of different species, as real granular materials
such as sand, gravel or seeds are rarely composed of identical
particles.

Starting with Jenkins and Mancini
\cite{jenkins+mancini87,jenkins+mancini89} \emph{binary} mixtures and
in particular their kinetic temperature and transport properties
received considerable interest \cite{garzo+dufty99,lu+liu00,
  barrat+trizac02,dahl+hrenya02,pagnani+marconi02,alam+luding03,
  galvin+dahl05, alam+luding05,garzo+montanero07}. These studies
confirmed that equipartition of energy is indeed violated in granular
binary mixtures, an observation that was first made in experiments by
Losert \textit{et. al.} \cite{losert+cooper99}. \emph{Polydisperse}
granular mixtures, i.e., mixtures composed of more than two types of
particles were studied much less \cite{zamankha95,dahl+clelland02,
  iddir+arastoopour05,lambiotte+brennig05,garzo+dufty07,
zhi-yuan+duan-ming08}
although they are closer to realistic systems. In particular, Dahl
\textit{et. al.}  \cite{dahl+clelland02} and Zhi-Yuan \textit{et. al.}
\cite{zhi-yuan+duan-ming08} simulated mixtures of particles with a
distribution of sizes, and Lambiotte \textit{et. al.}
\cite{lambiotte+brennig05} discuss mixtures of Maxwell molecules with
varying coefficients of restitution.

Out of the many fascinating phenomena inherent to granular mixtures
and the observables that are necessary to understand them, we will
focus on the partitioning of energy and how it evolves in time,
both in the homogeneous cooling state (\textsc{HCS}) and in
homogeneously driven systems. Even though, in this paper we will first
develop the machinery to deal with an arbitrary number, $X$, of
species, we will eventually go one step further and consider highly
polydisperse systems, where no two particles are alike but instead
possess properties that are drawn from continuous probability
distributions.

In the following three sections we give a short introduction to the
model and methods we use. In section \ref{sec:highly} we investigate
the temperature in a highly polydisperse system, characterized by a
continuous distribution of particle sizes. We finish with a brief
conclusion and delegate all technical material to the appendices.

\section{Model and observables}
\label{sec:model}

In order to model a polydisperse granular gas, we consider mixtures of
$X$ different species of smooth inelastic hard spheres. Each species
$\alpha = 1,2,\ldots,X$ consists of $N_{\alpha}\to\infty$ identical
particles, such that the concentrations $x_{\alpha} := N_{\alpha}/N$
($N = \sum_{\alpha}N_{\alpha}$) as well as the density $n = N/V$
remain finite as $N_{\alpha}\to\infty$. Collisions between particles
are assumed to be instantaneous and the particles move freely between
collisions. Because of the vanishing collision time collisions of more
than two particles can be neglected, i.e. the dynamics is determined
by two particle collisions. The inelasticity is described by a
velocity independent coefficient of normal restitution,
$\epsilon_{\alpha\beta} \in [0,1]$, which may depend on the pair of
species $\alpha,\beta=1,2,\ldots, X$ that the colliding particles
belong to:
\begin{equation}
  \label{eq:rest}
  \uvec n \cdot \vec v'_{12} 
  = -\epsilon_{\alpha\beta} \uvec n \cdot \vec v_{12},
\end{equation}
where $\vec v_{12} = \vec v_1 - \vec v_2$ is the relative velocity of
the colliding particles at contact before the collision and $\vec
v_{12}'$ the corresponding quantity after the collision.  The unit
vector $\uvec n$ points from the center of particle 1 to the center of
particle 2. Apart from the mutual coefficient of restitution
$\epsilon_{\alpha\beta}$, the species may also differ in mass
$m_{\alpha}$ and in size (radius) $a_{\alpha}$.

 The collision law [eq.~\eqref{eq:rest}] together with conservation of
 momentum determines the postcollisional velocities $\vec v'_1$ and
 $\vec v'_2$ uniquely in terms of the precollisional ones ($\vec v_1$,
 $\vec v_2$):
\begin{equation}
  \label{eq:vel}
  \begin{aligned}
    \vec v'_1 &=& \vec v_1 - \frac{m_2}{m_1+m_2}(1+\epsilon_{12})
    (\uvec n \cdot \vec v_{12})\uvec n, 	\\ 
    \vec v'_2 &=& \vec v_2 + \frac{m_1}{m_1+m_2}(1+\epsilon_{12})
    (\uvec n \cdot \vec v_{12})\uvec n
  \end{aligned}
\end{equation}
As we consider smooth
spheres the tangential component of the relative velocity ($\vec
v_{12} \times \uvec n$) remains unaffected.

Due to the inelasticity, the particles suffer an energy loss during
collision, i.e. the gas will cool down. To compensate for this energy
loss, one can provide the system with external energy. We will
restrict ourselves to volume driving \cite{williams+mackintosh96}:
With a given frequency $f_{\text{dr}}$ random kicks 
\begin{equation}
  \label{eq:drive}
  \vec p_i\to\vec p_i + p_{\text{dr}}\vec\xi_i(t)
\end{equation}
are applied to all particles individually ($\vec p_i\equiv m_i\vec
v_i$). The strength of the kicks is controlled by $p_{\text{dr}}$
while the components of $\vec\xi_i$ are drawn form a white noise
source: $\overline{\xi_i^a} = 0$ and $\overline{\xi_i^a(t)\xi_j^b(t')}
= \delta_{ij}\delta^{ab}\delta(t - t')$. The time between two driving
events is taken to be small compared to the time scale on which the
gas would cool down without energy supply.

When considering $X$-component mixtures, the driving strength
$p_{\text{dr}}$ may in general be a function of the particle species
$p_{\text{dr}}\equiv p_{\text{dr}}^{\alpha}$. There are several
experimental methods (both in $D=2$ and $D=3$) that one can hope to
describe approximately by volume driving: Shaking on a rough plate
\cite{prevost+egolf02}, electrostatic
\cite{aranson+olafsen02,kohlstedt+snezhko05} or magnetic
\cite{kohlstedt+snezhko05,maass+isert08} excitation, fluidisation by
air \cite{ohja+lemieux04,abate+durian05} or water
\cite{schroeter+goldman05}. As it is not obvious how to best describe
the driving of all these experiments theoretically, we propose the
following three simple mechanisms:
\begin{enumerate}
  \renewcommand{\theenumi}{\roman{enumi}}
  \itemsep=0em
\item force controlled driving, assuming that all particles experience
  the same force ($p_{\text{dr}}^{\alpha}\equiv p_{\text{dr}}$),
\item velocity controlled driving, assuming that all particles get
  velocity kicks of the same magnitude ($p_{\text{dr}}^{\alpha}\propto
  m_{\alpha}$) and
\item energy controlled driving, supplying every species on average
  with the same energy ($p_{\text{dr}}^{\alpha}\propto
  m_{\alpha}^{1/2}$).
\end{enumerate} 
The first two mechanisms combined with an additional viscous drag
force $\propto\eta\vec v$ are also discussed in the context of binary
mixtures by Pagnani \textit{et. al.}  \cite{pagnani+marconi02}. Our
hope is that the results discussed below may help to clarify the
experimental conditions.

The basic quantity of interest is the one-particle velocity
distribution, $f_\alpha(\vec v)\mathrm{d}^Dv$, of species $\alpha$
which is related to the one-particle distribution $f_{\alpha}(\vec r,
\vec v)\mathrm{d}^Dr\mathrm{d}^Dv$ by $f_{\alpha}(\vec v) = \int
f_{\alpha}(\vec r, \vec v)\mathrm{d}^Dr$. As an example, consider
species that differ in mass, so that the one-particle
velocity distribution is explicitly given by
\begin{equation}
  f_\alpha(\vec v)\mathrm{d}^Dv =\sum_i^N \delta_{m_i,m_{\alpha}} 
  \langle\delta(\vec v - \vec v_i)\rangle \,\mathrm{d}^Dv,\notag
\end{equation}
where the angular brackets
$\langle \cdot \rangle$ denote the average over the $N$-particle
distribution function. It is normalized such that
\begin{equation}
  \int\mathrm{d}^Dv\,f_\alpha(\vec v) = N_{\alpha}
  \quad \mbox{and} \quad
  \sum_{\alpha}\int\mathrm{d}^Dv\,f_\alpha(\vec v)=N.\notag
\end{equation}
The partial granular temperature for species $\alpha$ in $D$ space
dimensions is defined by
\begin{equation}
  \label{eq:Ta}
  \frac{D}{2}T_{\alpha}:=\frac{1}{N_{\alpha}}\sum_i
  \frac{m_{\alpha}}{2}\avr{v_i^2}\delta_{m_i,m_{\alpha}}
  = \frac{\int \mathrm{d}^Dv\,f_\alpha(\vec v)\frac{m_{\alpha}v^2}{2}}
  {\int\mathrm{d}^Dv\,f_\alpha(\vec v)}.
\end{equation}
The mean temperature, $\overline T=\sum_{\alpha}x_{\alpha}T_{\alpha}$,
is then just given by the mean kinetic energy
\begin{equation}
  \frac{D}{2} \overline T= \frac1N\sum_{\alpha}\int\mathrm{d}^Dv\,f_\alpha(\vec
  v)\frac{m_{\alpha}v^2}{2}
  =\frac{1}{N}\sum_i\frac{m_i}{2}\langle v_i^2 \rangle.\notag
\end{equation}
The above definitions are easily generalized to other species
characteristics, e.g. different size or different coefficients of
restitution: The indicator function, $\delta_{m_i,m_{\alpha}}$, just
has to be replaced by the corresponding one.

Our main emphasis in this paper are particles whose properties depend
on a continuous variable $\alpha\in\mathbb{R}$ that follows a
prescribed probability distribution $d\mu(\alpha)$, i.e.
\begin{equation}
  \sum_{\alpha}\frac{N_{\alpha}}{N} \to 
  \int\mathrm{d}\alpha\, x(\alpha)=\int\mathrm{d}\mu(\alpha).\notag
\end{equation}
The temperature becomes a continuous function $T_\alpha\to T(\alpha)$
whose mean and variance is given by 
\begin{equation}
  \label{eq:mean}
  \begin{aligned}
    \overline T= \int T(\alpha)\mathrm{d}\mu(\alpha)\, ,\quad 
    \Delta T= \overline {T^2}-\overline T^2\\
    \mbox{with} \quad \overline{T^2} = \int T^2(\alpha)\mathrm{d}\mu(\alpha).
  \end{aligned}
\end{equation}  
In our example of a distribution of masses, $\alpha=m$ the one-particle
velocity distribution, $f(m,\vec v)\mathrm{d}^3v\,\mathrm{d}m $, is defined by
\begin{equation}
  f(m,\vec v) = \sum_i^N \delta(m_i-m) 
  \langle\delta(\vec v - \vec v_i)\rangle.\notag
\end{equation}

\section{Analytical theory}
\label{sec:analytical}

The time evolution of the temperatures is computed with the help of
the pseudo Liouville operator formalism. For details see
e.g. refs. \onlinecite{huthmann+zippelius97} and \onlinecite{aspelmeier_huthmann_zippelius_free_cooling}. In
this framework the time evolution of an observable $A$ is given by the
equation
\begin{equation}
  \frac{\mathrm{d}}{\mathrm{d}t} \langle A \rangle =
  \langle i \mathcal{L}A\rangle,\notag
\end{equation}
where $i\mathcal{L}$ denotes the pseudo Liouville operator.

The pseudo Liouville operator for the driven hard sphere gas consists
of three terms. The term $i\mathcal{L}_0$ describes free
streaming which does not affect the temperature, the term
$i\mathcal{L}_H$ accounts for driving and 
$i\mathcal{L}_I$ for interactions between particles.
In a gas consisting of $X$ different species one obtains
\begin{equation}
  i\mathcal{L} = i\mathcal{L}_0 + i\mathcal{L}_H +
  \sum \limits_{\alpha=1}^X 
  \sum \limits_{\beta=1}^{\alpha} 
  i\mathcal{L}_{\alpha\beta},\notag
\end{equation}
where $i\mathcal{L}_{\alpha\beta}$ accounts for interactions between
particles of species $\alpha$ with particles of species $\beta$.  For
the evolution of the temperature of a particular species, only
interactions with participation of that species play a role;
collisions between particles of other species do not have a direct
influence. Given a discrete number, $X$, of different species, the
temperature of species $\alpha$, eq.~\eqref{eq:Ta}, develops in the
following way
\begin{subequations}
  \begin{equation}
    \label{eq:discrete}
    \frac{D}{2}\frac{\mathrm{d}}{\mathrm{d}t}T_{\alpha}
    = \langle i \mathcal{L}_H \overline {E}_{kin}(\alpha) \rangle 
    + \sum \limits_{\beta=1}^X \langle i\mathcal{L_{\alpha\beta}}
    \overline {E}_{kin}(\alpha) \rangle.
  \end{equation}
  Given a continuous distribution $d\mu(\alpha)$ of a parameter
  $\alpha$, one obtains
  \begin{equation}
    \label{eq:cont}
    \frac{D}{2}\frac{\mathrm{d}}{\mathrm{d}t} T(\alpha)
    = \langle i\mathcal{L}_H \overline {E}_{kin}(\alpha) \rangle 
    + \int \langle i \mathcal{L_{\alpha\beta}}\overline {E}_{kin}(\alpha)
    \rangle \mathrm{d}\mu(\beta).
  \end{equation} 
\end{subequations} 
At this point we would like to stress that the above equations hold
subject to arbitrary initial conditions $T_{\alpha}(t=0)$. The
\textit{a priori} assumption of a (quasi-)stationary state that is
required for some of the hydrodynamic theories is not needed here.

For a hard core potential the interaction terms
$i\mathcal{L}_{\alpha\beta}$ separate into a sum of two particle
interaction operators $i \mathcal{L}_{\alpha\beta} = \frac{1}{2}
\sum_{k,l} i \mathcal{T}_{\alpha\beta}^{(kl)}$ with one particle
belonging to species $\alpha$, the other one to species $\beta$. For
the operator $i\mathcal{T}_{\alpha\beta}^{(kl)}$ one obtains
\begin{equation}
  i\mathcal{T}_{\alpha\beta}^{(kl)} := 
  - (\vec v_{kl} \cdot \uvec n) 
  \Theta(-\vec v_{kl} \cdot \uvec n) \delta(r_{kl} - a_k -a_l)
  (b_{\alpha\beta}^{(kl)}-1),\notag
\end{equation}
where $b_{\alpha\beta}^{(kl)}$ is the operator replacing the
particles' velocities before collision by their values afterwards
according to equation \eqref{eq:vel}.

When calculating the phase space average, one has to take into account
the excluded volume effect which arises due to the fact that particles
cannot overlap. Consequently, the phase space element in $D$
dimensions is given by
\begin{equation}
  \mathrm{d}\Gamma = \prod_{i < j} \Theta(r_{ij}-a_i-a_j) 
  \prod_{k=1}^{N_1}\mathrm{d}^Dr_k\mathrm{d}^Dv_k \dots 
  \prod_{\ell=1}^{N_X} \mathrm{d}^Dr_{\ell}\mathrm{d}^Dv_{\ell}\notag
\end{equation}
with $r_{ij}$ the distance between particles $i$ an $j$.

We assume that the particles are uniformly distributed in space, that
the species are well mixed and that velocity correlations between
different particles can be neglected (molecular chaos
assumption). Under these premises the $N$-particle distribution
function $f_N(\{\vec r_i\}, \{\vec v_i\}, t)$ factorizes into a
product of $N$ single particle distribution functions $f(\vec r, \vec
v, t)$. In a monodisperse system, the single particle distribution
function can be written in rescaled form
\begin{equation}
  \label{eq:2}
  f(\vec r, \vec v, t) 
  \propto \frac{n}{T(t)^{D/2}}\tilde f(\vec v/\sqrt{T(t)})\notag
\end{equation}
both in the homogeneous cooling state as well as in the stationary
state of a driven system \cite{noije_ernst_velocity_distributions}.

Contrary to elastic gases, a Gaussian distribution is only an
approximate solution in the inelastic case. Deviations have been
studied extensively for driven and undriven monodisperse systems.
Investigations have shown that while a Gaussian approximation is quite
good in the range of typical velocities, high velocities are
overrepresented in granular gases
\cite{goldshtein+shapiro95,noije_ernst_velocity_distributions,brilliantov+poeschel00,ernst+brito03,poeschel+brilliantov06}. In
ref.~\onlinecite{garzo+dufty99} qualitatively similar deviations have
been found for freely cooling binary mixtures. The corrections have,
however, only little influence on the temperature and the cooling rate
\cite{noije_ernst_velocity_distributions}. Thus, we make a Gaussian
ansatz for the velocity distribution of a single species $\alpha$ with
temperature $T_{\alpha}$. The $N$-particle distribution for a mixture
with $X$ components then follows:
\begin{equation}
  \label{eq:distribution}
  f_N(\{\vec r_i\},\{\vec v_i\}, t)
  \propto  
  \prod_{i=1}^{N_1}e^{-\frac{m_1\vec v_i^2}{2 T_1(t)}}\cdots
  \prod_{j=1}^{N_X}e^{-\frac{m_X\vec v_j^2}{2 T_X(t)}}.
\end{equation}

In undriven systems, the \textsc{HCS} is maintained only for a certain
time until velocity correlations develop and clusters form because of
the system's instability against density fluctuations
\cite{hopkins+louge91,mcnamara93,goldhirsch+zanetti93}. In inelastic
mixtures cluster formation is additionally accompanied by the onset of
segregation \cite{cattuto+marconi04,galvin+dahl05}. Therefore our
results will in this case be limited to the initial development.

Using the distribution function, eq.~\eqref{eq:distribution}, evaluation 
of the term 
$\langle i \mathcal{L}_{\alpha\beta} \overline {E}_{kin}(\alpha)\rangle$
yields (cf. appendix~\ref{sec:calc-mixed-term}):
\begin{widetext}
  \begin{equation}
    \label{eq:eval}
    \langle i \mathcal{L}_{\alpha\beta} \overline {E}_{kin}(\alpha)\rangle
    = -2x_{\beta} \mu_{\alpha \beta} 
    G_{\alpha \beta} 
    \sqrt{\frac{T_{\alpha}m_{\beta} + T_{\beta}m_{\alpha}}{2m_{\beta}}}
    \left[
      \frac{A_{\alpha \beta}}{m_{\alpha}}T_{\alpha}
      - \frac{B_{\alpha \beta}}{m_{\beta}}(T_{\beta}-T_{\alpha})
    \right]
  \end{equation}
\end{widetext}
with the reduced mass $\mu_{\alpha \beta} :=
m_{\alpha}m_{\beta}/(m_{\alpha}+m_{\beta})$. The other constants are
given by
\begin{equation}
  A_{\alpha \beta} := \frac{1-\epsilon_{\alpha \beta}^2}{4},\quad
  B_{\alpha \beta} := \frac{1}{4} \frac{(1+\epsilon_{\alpha \beta})^2}
  {1 + \frac{m_{\alpha}}{m_{\beta}}}\notag
\end{equation} 
and
\begin{equation}
  \begin{aligned}
    G_{\alpha \beta} &:= 4(a_{\alpha}+a_{\beta})n 
    \sqrt{\frac{\pi}{m_{\alpha}}}\chi_{\alpha \beta}
    &&\text{for} \ D=2, \\
    G_{\alpha \beta} &:= 8(a_{\alpha} + a_{\beta})^2 n 
    \sqrt{\frac{\pi}{m_{\alpha}}}\chi_{\alpha \beta}
    &&\text{for} \ D=3
  \end{aligned}\notag
\end{equation}
where $\chi_{\alpha \beta}$ is the value of the pair correlation
function $g_{\alpha\beta}(r)$ at contact. In the following, we will
use the approximation $\chi_{ra}=1$ which is well justified for dilute
systems.

The terms in equation \eqref{eq:eval} have a direct physical
interpretation: The factor before the square brackets defines an
effective collision frequency $\omega_{\alpha\beta}$ of particles
coming from possibly different species with different
temperatures. The first term inside the brackets accounts for the
dissipation in collisions between $\alpha$ and $\beta$ particles while
the second term describes the heat flux between species with different
temperatures which tends to equalize the two temperatures. This term
is the only one present in mixtures of elastically colliding
particles, where it ensures equipartition. The difference to the
elastic cases consists in the dissipative terms. As the cooling rates
$\propto G_{\alpha\beta}A_{\alpha\beta}$ are in general different for
each species and are completely independent from the rate of energy
exchange $\propto G_{\alpha\beta}B_{\alpha\beta}$ they constantly
drive the system away from equipartition.  The new quasi-stationary
state is then no longer characterized by equipartition but by {\it
  equal cooling rates} $\dot T_{\alpha}/T_{\alpha} = \dot
T_{\beta}/T_{\beta}$ \cite{garzo+dufty99}. A related interpretation
has been given before by Alam and Luding
\cite{alam+luding05}. Moreover it is also apparent that driving the
system will in general not be sufficient to restore equipartition as
was first shown by Barrat and Trizac \cite{barrat+trizac02}. For the
special case of an undriven system that already reached its
quasi-stationary state, equation~\eqref{eq:discrete} is equivalent to
equation~(2.4) of ref.~\onlinecite{garzo+dufty07}a.

The driving power $H_{\alpha}$ which was formally written as
$H_{\alpha} = \langle i\mathcal{L}_H\overline{E}_{kin}(\alpha)\rangle$
in equations \eqref{eq:discrete} and \eqref{eq:cont} can be more
easily calculated directly form the definition [eq.~\eqref{eq:drive}]:
\begin{equation}
  \label{eq:1}
  H_{\alpha} = f_{\text{dr}}(p_{\text{dr}}^{\alpha})^2/2m_{\alpha}.\notag
\end{equation}
In particular we get for (i) force controlled driving
$H_{\alpha}^{\text{fc}} \propto 1/m_{\alpha}$, (ii) velocity
controlled driving $H_{\alpha}^{\text{vc}} \propto m_{\alpha}$, and
(iii) constant energy input $H_{\alpha}^{\text{ec}}$ independent of
$m_{\alpha}$.

\section{Simulations}

In order to test our analytical theory we performed complementary
computer simulations based on an event-driven (\textsc{ED}) algorithm
\cite{lubachevsky91}. Although our code can easily handle up to
$10^6$ particles, we usually found $10^4$ particles per species
sufficient for the measurements reported here. Because of the
extremely low densities used in this paper, we hardly ever need to
take care of the inelastic collapse occurring in
\textsc{ED}-simulations. If necessary we use the method of
ref.~\onlinecite{fiege09} to avoid inelastic collapse.

For monodisperse systems, the minimal cluster size $L_c$ can be
derived from a hydrodynamic stability analysis
\cite{mcnamara93,garzo05}. To keep our systems from clustering, we
chose a system size $L\lesssim L_c/6$. Although $L_c$ will certainly
be somewhat different for polydisperse systems, we found no
indications for clustering or segregation in our simulations.

As mentioned above, our simulations include volume driven systems.  In
this context it is necessary that the simulation process takes the
conservation of momentum into account. To do so, a driving event
always concerns two particles at the same time \footnote{A similar
  scheme is employed in dissipative particle dynamics (see
  e.g. \cite{espanol+warren95}). For the far reaching consequences the
  choice of driving mechanism can have, see
  \cite{fiege09,aspelmeier09}.}. One of these particles, say particle
1, is chosen at random. The neighborhood of this particle is examined
to find the particle, $i$, closest to the first one. Particles 1 and
$i$ are then kicked at the same time $t$. While a momentum increment
$p_{\text{dr}}\vec\xi(t)$ [see eq.~\eqref{eq:drive}] is \emph{added}
to particle 1, it is \emph{subtracted} from particle $i$, i.e.
\begin{equation}
  \label{eq:4}
  \begin{aligned}
    \vec p_1&\to\vec p_1 + p_{\text{dr}}\vec\xi\\
    \vec p_i&\to\vec p_i - p_{\text{dr}}\vec\xi
  \end{aligned}.\notag
\end{equation}
In that way momentum is conserved on length scales $\ell$ of a mean
particle separation, i.e., $\ell\propto n^{-1/D}$.

The simulations were performed in two steps. Initially the particles
were placed on a grid and random velocities drawn from a Gaussian
distribution were assigned to the particles. In the first half of the
simulation, all coefficients of restitution were set to unity and the
elastic mixture was simulated for about 120 collisions per particle to
generate a well mixed state. In the next step the desired
inelasticities were switched on and the temperatures were recorded
until reliable estimates for the stationary values of the observables
could be obtained. For the driven systems we chose the driving
frequency $f_{\text{dr}}$ to be approximately the same as the
collision frequency at the desired stationary temperature
$T_{\infty}$. As a compromise between computational efficiency and the
desire to reduce temperature fluctuations due to rare but strong
driving events this choice of driving frequency was also found
satisfactory by Bizon \textit{et. al.} \cite{bizon+shattuck99}.
 
\section{Highly polydisperse systems}
\label{sec:highly}

Many real granular systems are highly polydisperse with no single
particle being identical in shape and size to another one. To account
for a high degree of polydispersity we generalize the considerations
for polydisperse mixtures to mixtures containing "infinitely" many
species. In principle, a variety of scenarios can be thought of and
treated within our analytical approach. Here, we will restrict
ourselves to the relatively simple case where the particles' radius is
uniformly distributed in a range $[R_1,R_2]$; the particles all have
the same mass density $\rho$ and all restitution coefficients are
equal $\epsilon_{\alpha\beta}\equiv\epsilon$. We furthermore choose
units such that $\rho = 1$.

The following questions are of particular interest. Is there a
stationary temperature profile, $T(a)$, if the system is driven?  If
so, how does this function reflect the properties of the distribution
of radii? How does the forcing mechanism affect the stationary
temperature profile? How does the system cool freely if undriven?

Combining equations \eqref{eq:cont} and \eqref{eq:eval} leads to the
following integro-differential equation for the temperature of species
with radius $a$:
\begin{equation}
  \label{eq:integrodiff}
  \frac{D}{2}\frac{\mathrm{d}}{\mathrm{d}t}T(a) = H(a) + \mathsf F[T](a)
\end{equation}
where the nonlinear integral operator $\mathsf F$ is given (in $D=3$)
by
\begin{widetext}
  \begin{equation}
    \mathsf F[T](a) := \frac{n\sqrt{6}}{R_2-R_1} \int
    \limits_{R_1}^{R_2}\mathrm{d}r\chi_{ra}\frac{r^3(a+r)^2}{r^3+a^3}
    \sqrt{\frac{T(a)}{a^3} + \frac{T(r)}{r^3}}
    \Bigg\{(\epsilon^2-1)T(a) + (1+\epsilon)^2 \frac{a^3}{a^3+r^3}
    \left[T(r)-T(a)\right]\Bigg\}.\notag
  \end{equation}
\end{widetext}

When the system is driven constantly in time, we expect a stationary
temperature profile $T_{\infty}(a) = T(a, t\to\infty)$, to develop. If
this is correct, it should be given as the asymptotic solution of
equation \eqref{eq:integrodiff} with the left hand side set to zero:
\begin{equation}
  \label{eq:integral_temp}
  \mathsf F[T_{\infty}](a) = -H(a)
\end{equation}
In general $T_{\infty}$ depends not only on $a$ but also on the two
parameters $R_1,R_2$ of the distribution of radii.  By scaling all
radii with $R_1$, one observes that (up to a scale factor)
$T_{\infty}$ depends only on the ratios $a^*=a/R_1$ and $R = R_2/R_1$,
but not on the absolute values. Alternatively, we choose $a^*$ and the
relative width of the distribution $\Delta=2(R_2-R_1)/(R_2+R_1)$ as
independent variables: $T_{\infty}=T_{\infty}(a^*,\Delta)$.

We solved the above nonlinear integral equation
\eqref{eq:integral_temp} numerically by applying Banach's fixed point
iteration (for details see appendix~\ref{sec:solv-integr-equat}). We
always found a solution, confirming that a stationary temperature
profile is indeed reached for asymptotically long times.

\begin{figure}
  \includegraphics{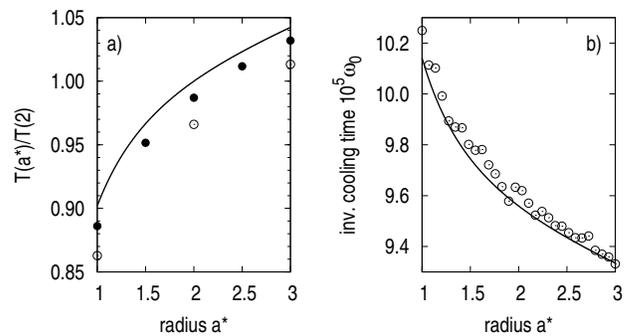}
  \caption{a) The stationary temperatures [eqs.~\eqref{eq:discrete} \&
    \eqref{eq:cont}] in a three component (open disks) and five
    component (filled disks) mixtures compared to those of a highly
    polydisperse mixture for energy controlled driving
    $H^{\text{ec}}=10^{-3}$ at density $n=5\times10^{-4}$ and
    coefficient of restitution $\epsilon = 0.9$. b) Inverse cooling
    time $\omega_0$ in a two dimensional system for a uniform size
    distribution of width $R = 3$, coefficient of restitution
    $\epsilon = 0.9$ and density $n = 2\times10^{-4}$. The symbols
    denote simulation results for $X=30$ species each with $10^4$
    particles, while the solid line is the solution of
    eq.~\eqref{eq:integral_omega}.}
  \label{fig:sim1}
\end{figure}

Independently, we performed event driven simulations and measured all
the partial temperatures $T(a,t)$.  The amount of simulation time
needed for sufficiently good statistics quickly rises with the number
of species. To this end, we checked if a \emph{highly polydisperse}
system can be approximated by a \emph{polydisperse} mixture with many
species such that there is still a considerable number of particles
for each species. Considering equation \eqref{eq:discrete} for
increasing numbers of species we found that the temperatures
considered in this paper rapidly converge. Figure~\ref{fig:sim1}(a)
shows how mixtures of respectively three and five species compare to
the result for a continuous distribution. From these results we
conclude that considering $X = 20$--$30$ species for the simulations
should yield results practically indistinguishable from the highly
polydisperse case.

\begin{figure}
  \includegraphics{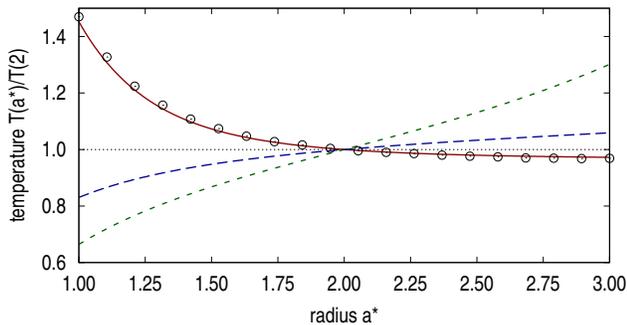}
  \caption{(Color online) The stationary temperature in a three
    dimensional driven system for a size distribution with $R_2=3R_1$,
    a coefficient of restitution $\epsilon = 0.9$ at a density
    $n=2\times10^{-4}$. Force controlled driving $H^{\text{fc}}(a) =
    1.875\times10^{-3}/m(a)$ (solid, red), energy controlled driving
    $H^{\text{ec}} = 1.875\times10^{-3}$ (long dashed, blue) and
    velocity controlled driving $H^{\text{vc}}(a) =
    1.875\times10^{-3}m(a)$ (short dashed, green). For the simulation
    data (symbols) a system of 20 different particle species with
    $10^4$ particles each was used. }
  \label{fig:sim4}
\end{figure}

In \textsc{FIG}.~\ref{fig:sim4} we show the stationary temperature
$T_{\infty}(a^*,\Delta)$ as a function of particle radius $a^*$ for
the three driving mechanisms proposed in section \ref{sec:model}.  The
rough trends can be understood from the following qualitative
arguments.  Force controlled driving $H^{\text{fc}}(a^*)\propto
1/m(a^*)\propto a^{*-3}$ is dominant for small particles so that one
expects the partial temperatures, $T_{\infty}(a^*,\Delta)$, to
decrease with increasing size $a^*$. This is indeed born out by the
solution of the integral equation \eqref{eq:integral_temp} and
supported by simulations, which are seen to agree well with the
theoretical result. Velocity controlled driving $H^{\text{vc}}\propto
m(a^*)$ is dominant for large particles so that we expect the partial
temperatures to increase with increasing size of the particles, as is
indeed observed in \textsc{FIG}.~\ref{fig:sim4}.  Finally, for the
energy controlled mechanism, $H^{\text{ec}}(a^*)\equiv H$, is
independent of the particle size, nevertheless
$T_{\infty}(a^*,\Delta)$ depends weakly on $a^*$. One has to keep in
mind that all the species interact and that this will lead to
nontrivial conditions of stationarity as in the binary case. These
effects are responsible for the precise functional form of the
temperature profile which goes beyond the simple rough trend for
all three driving mechanisms. The same trends for force controlled
versus velocity controlled driving have been found by Pagnani
\textit{et. al.} \cite{pagnani+marconi02} in the case of binary
mixtures.

Abate and Durian \cite{abate+durian05} discuss several systems that,
although they are comprised of only two to five particles come close
to our definition of highly polydisperse systems in that no two
particles are alike. Two spheres of different sizes show a marked
increase in the temperature ratio with increasing size ratio. This
would roughly correspond to our results for velocity controlled
driving but the authors of ref.~\onlinecite{abate+durian05} observed a
complicated two particle interaction. Moreover, they considered a
system with five different spheres of the same size but different
densities. Based on the results from binary mixtures one infers that
the effects of different masses is much stronger than that of
different sizes. If this reasoning is valid the weak dependence of the
temperature on the mass would correpond to energy controlled driving
in the present paper.

\begin{figure}
    \includegraphics{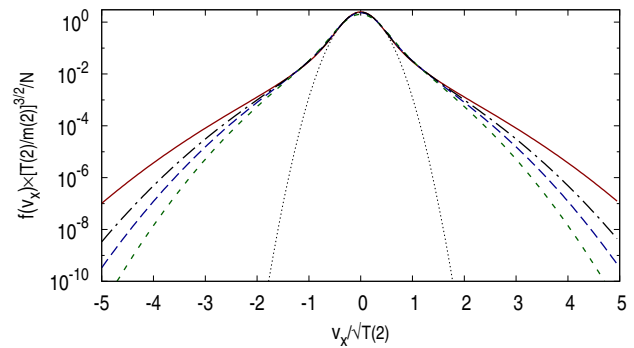}
    \caption{(Color online) Stationary velocity distribution
      [eq.~\eqref{eq:pv_total}] in a three dimensional driven system;
      parameters and symbols as in \textsc{FIG}.~\ref{fig:sim4}; for
      comparison the velocity distribution of an elastically
      colliding, molecular gas (dashed-dotted, black) and a gaussian
      fit to the central part of the distribution (thin dotted line)
      are also shown.}
  \label{fig:pv}
\end{figure}

Within our approximation scheme, the partial temperatures (i.e., the
temperature profile), $T_{\infty}(a^*)$, determine the one-particle
velocity distribution according to
\begin{equation}
  f(a,\vec v) = \frac{N}{R_2 - R_1}\left[\frac{m(a)}{2\pi T_\infty(a)}\right]
  ^{D/2}e^{-m(a)\vec v^2/2 T_\infty(a)}.\notag
\end{equation}
The total velocity distribution, $f(\vec v)\mathrm{d}^3v$ is thus given by
\begin{equation}
  \label{eq:pv_total}
  f(\vec v) = \int\limits_{R_1}^{R_2}\!\mathrm{d}a f(a,\vec v).
\end{equation}
This function is in general not Gaussian, not even for an elastic
molecular gas with many different species. In
\textsc{FIG}.~\ref{fig:pv} we show the total velocity distribution as
given by equation \eqref{eq:pv_total}. The elastic system
(dashed-dotted) is compared to the inelastic gas with different
driving mechanisms. In comparison to the molecular gas the tails of
the velocity distribution can either be overpopulated, as observed for
force controlled driving (solid line), or underpopulated for energy
(long dashed) or velocity controlled (short dashed)
driving.

\begin{figure}
    \includegraphics{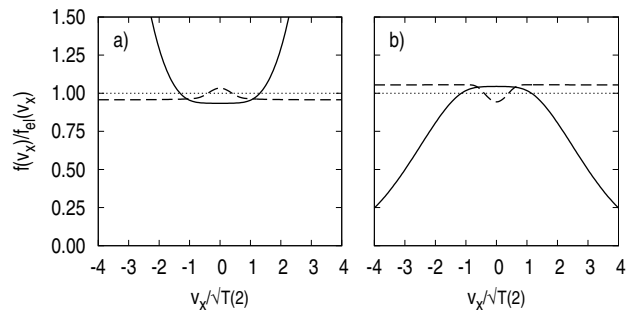}
    \caption{Stationary velocity distribution relative to the elastic
      gas and separately for the two halves of smaller (full line) and
      larger (dashed line) particles; parameters as in
      \textsc{FIG}.~\ref{fig:sim4}; a) force controlled driving, b)
      energy controlled driving.}
  \label{fig:pvsl}
\end{figure}

To clearly see the difference to the elastic case, we plot in 
\textsc{FIG}.~\ref{fig:pvsl} the velocity distribution relative to the
elastic gas. We furthermore separate the particles into two halves,
one with the smaller and one with the larger particles. The strongest
deviations are clearly in the tails and solely due to the small
particles. The velocity distribution of the large particles has
almost the same form as in the elastic gas, except for very small
velocities. Force and energy controlled driving are almost mirror
images of each other --- even for the detailed structures at small
velocities.    

\begin{figure}
  \includegraphics{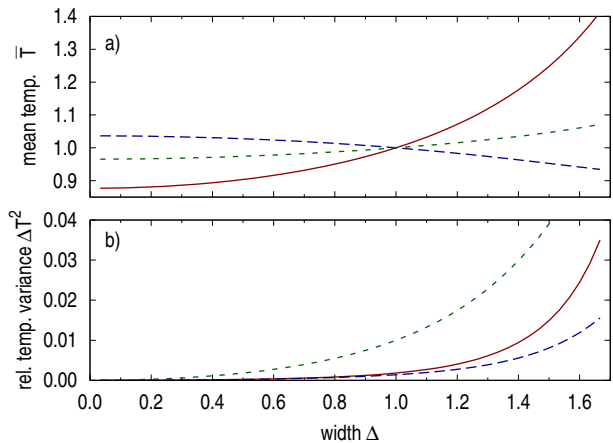}
  \caption{(Color online) a) The mean temperature $\overline T$
    [eq.~\eqref{eq:mean}] and b) the relative temperature variance
    $\Delta T^2 = \overline{T^2}/\overline T^2 - 1$ as a function of
    the relative width of the size distribution $\Delta$ for the three
    driving mechanisms $H^{\text{fc}}(a) = 10^{-2}/m(a)$ (solid, red),
    $H^{\text{ec}}(a) = 10^{-2}$ (long dashed, blue) and $H^{\text{vc}}(a) =
    10^{-2}m(a)$ (short dashed, green), coefficient of restitution $\epsilon
    = 0.9$ and density $n = 5\times10^{-4}$. The mean temperatures in
    (a) are rescaled such that they agree for a relative width of
    $\Delta = 1$.}
  \label{fig:stationaer_int_werte}
\end{figure}

How does the temperature profile, $T_{\infty}(a^*,\Delta)$, reflect
the prescribed distribution of radii? The latter is characterized by a
single parameter, the relative width $\Delta$, which can take values
$0\leq\Delta \leq 2$. In \textsc{FIG}.~\ref{fig:stationaer_int_werte}
we show the mean temperature $\overline T$ and the temperature
variance $\Delta T^2 := \overline{T^2} - \overline T^2$ (see
eq.~\eqref{eq:mean}) as a function of $\Delta$.  In
\textsc{FIG}.~\ref{fig:stationaer_int_werte}(a) we show the mean
temperature $\overline T(\Delta)/\overline T(1)$, scaled such that
they coincide at $\Delta=1$. Surprisingly the dependence is
nonmomotonic for different driving mechanisms: whereas $\overline
T(\Delta)$ increases with $\Delta$ for force and velocity controlled
driving, $\overline{T}(\Delta)$ decreases with $\Delta$ for energy
controlled driving. The strongest variation is observed for force
controlled driving.  The corresponding variance of the temperature
profile [\textsc{FIG}.~\ref{fig:stationaer_int_werte}(b)] increases
trivially with $\Delta$. The variance for velocity controlled driving
is almost an order of magnitude larger than for the other two driving
mechanisms.

We next consider the freely cooling case ($H(a)\equiv0$)
\cite{lambiotte+brennig05,garzo+dufty07}. 
We expect Haff's law \cite{haff83} to hold also for $T(r,t)$ and hence
make the ansatz
\begin{equation}
  \label{eq:3}
  T(a,t)\propto\omega_0(a)^{-2}t^{-2}\notag
\end{equation}
for large times. This leads to an integral equation for the inverse
cooling time $\omega_0$
\begin{equation}
  \label{eq:integral_omega}
  -D\omega_0^{-2}(a) = \mathsf F[\omega_0^{-2}](a)
\end{equation}
Similarly to $T_{\infty}$, the decay rate $\omega_0$ depends only on
$a^*=a/R_1$ and $R = R_2/R_1$, or alternatively $\Delta$, but not on
the absolute values: $\omega_0=\omega_0(a^*,\Delta)$. The above
integral equation is solved numerically by subsequently applying
Banach's fixed point iteration and Newton's method (for details see
appendix~\ref{sec:solv-integr-equat}).  To extract
$\omega_0(a^*,\Delta)$ from the simulations, we performed simulations
with $X=30$ species, measured all partial temperatures and fitted them
to Haff's law. The resulting decay rates are plotted in
\textsc{FIG}.~\ref{fig:sim1}(b). The rate is seen to be a
monotonically decreasing function of $a^*$, however the dependence is
weak. Since the coefficient of restitution is the same for all
particles, this is a pure size effect, implying that smaller particles
relax faster than larger ones. The simulation data are seen to agree
well with the theoretical results, but show a considerable
scatter. This is most likely due to the difficulty in fitting the data
to Haff's law, given the uncertainty in time scale, when the
asymptotic decay applies.

\section{Conclusion}
\label{sec:conclusion}

We examined the partitioning of energy in highly polydisperse mixtures
of smooth hard spheres. The properties of the particles, such as mass,
radius or coefficient of restitution, are chosen from a continuous
distribution giving rise to a corresponding continuous temperature
profile. The latter has been computed approximately, generalizing
previous approaches of mixtures with several species. The analytical
theory leads to a nonlinear integro-differential equation for the time
dependent temperature profile, which has been solved numerically.

Our results are supported by event driven simulations for mixtures
with $X=20-30$ species. The good agreement between \textsc{ED}
simulations and the analytical theory indicates that the assumptions
of homogeneity and molecular chaos that are fundamental to the theory
are indeed observed in the simulated system. The direct simulation
monte carlo (\textsc{DSMC}) method \cite{bird76}, otherwise well
suited for dilute (granular) gases (see
e.g. \cite{brey+ruiz-montero96,brey+dufty97,montanero+santos00,brey+ruiz-montero05}),
would not have been able to show this as it ensures both homogeneity
and molecular chaos by construction.

As a specific example we have studied a {\it uniform} size
distribution in detail. We showed that a highly polydisperse mixture
still obeys Haff's law during free cooling. The distribution of sizes
gives rise to a nonuniform distribution of cooling rates, such that
the smaller particles are cooling faster.

A driven system relaxes to a stationary temperature profile which is
in general {\it nonuniform}. Depending on the driving mechanism, its
weight can be predominantly at small or large particles. If the
particles are driven by a constant force, then the smaller particles
are hotter. If the driving process supplies either a constant energy
or velocity, then the larger particles have a higher temperature. The
temperature profile reflects the distribution of radii, characterized
by the relative width $\Delta$. The variance of the temperature
increases with $\Delta$, as one would expect, whereas the mean
temperature can either increase (constant force driving) or decrease
with $\Delta$ (constant energy supply).

This strong dependence on the driving mechanism is also observed in
the velocity distributions. For a polydisperse system, these are in
general weighted sums of all partial distributions and hence in
general not Gaussian, even if the partial distributions are Gaussian
like in an elastic gas. The velocity distribution in an inelastic,
driven gas can have either {\it overpopulated} or {\it underpopulated} tails,
as compared to the molecular gas. Furthermore, the effects are
dominated by the small particles.

\begin{acknowledgments}
We thank M. Sperl for carefully reading the manuscript and M. Schr\"oter
for valuable suggestions.
\end{acknowledgments}

\appendix

\section{Calculation of the mixed term}
\label{sec:calc-mixed-term}

We will only show the calculations for the mixed term
$\avr{i\Lv_{12}T_1}_t$ for $D=2$. The calculations in $D=3$ are very
similar although slightly longer and the single species terms have
already been calculated (see, e.g.,
\citep{aspelmeier_huthmann_zippelius_free_cooling}). The first steps
are straight forward
\begin{equation}
  \label{eq:13}
  \begin{aligned}
    \avr{i\Lv_{12}T_1} &= \avr{\frac12\sum_{i,j}i\mathcal T^{(ij)}_{12}T_1}\\
    &= \frac{1}{2N_1}\sum_{i,j}
    \avr{i\mathcal T^{(ij)}_{12}\frac{m_1}{2}\sum_{k=1}^{N_1}\vec v_k^2}\\
    &= \frac{N_1N_2}{N_1}\avr{i\mathcal T^{(12)}_{12}\frac{m_1}{2}\vec v_1^2}
  \end{aligned}\notag
\end{equation}
where we used the molecular chaos assumption to reduce the average
over all possible pairs of colliding spheres to a sum of $2N_1N_2$
times the average result of a single colliding pair.

Now we introduce two partitions of unity ($\int
d^2R_1d^2R_2\delta(\vec R_1 - \vec r_1)\delta(\vec R_2 - \vec r_2)$),
i.e.
\begin{widetext}
  \begin{equation}
    \label{eq:14}
    \avr{i\Lv_{12}T_1} = N_2\frac{\int \mathrm{d}\Gamma
	\int \mathrm{d}^2R_1\mathrm{d}^2R_2
      \delta(\vec R_1 - \vec r_1)\delta(\vec R_2 - \vec r_2)
      f_N(\{\vec v_i\}, t)i\mathcal T^{(12)}_{12}\frac{m_1}{2}\vec v_1^2}
    {\int \mathrm{d}\Gamma f_N(\{\vec v_i\}, t)}\notag
  \end{equation}
  identifying the pair correlation function $g_{12}(R_{12})/V^2 =
  \avr{\delta(\vec R_1 - \vec r_1)\delta(\vec R_2 - \vec r_2)}_t$ in
  this expression yields
  \begin{equation}
    \label{eq:15}
    \avr{i\Lv_{12}T_1} = \frac{N_2}{V^2}\frac{\int 
	\mathrm{d}^2R_1\mathrm{d}^2R_2
      \int\prod_j
	\mathrm{d}^2v_jg_{12}(R_{12})f_N(\{\vec v_i\}, t)
      i\mathcal T^{(12)}_{12}\frac{m_1}{2}\vec v_1^2}
    {\int\prod_j\mathrm{d}^2v_jf_N(\{\vec v_i\}, t)}\notag
  \end{equation}
  Substituting $\vec R_{12} = \vec R_1 - \vec R_2$ for $\vec R_1$ the
  other spatial integration is trivial as are all the velocity
  integrals in the denominator and those for $j > 2$ in the numerator:
  \begin{equation}
    \label{eq:16}
    \avr{i\Lv_{12}T_1} 
    = -x_2n\left[\frac{m_1}{2\pi T_1(t)}\right]
    \left[\frac{m_2}{2\pi T_2(t)}\right]
    \int \mathrm{d}^2R_{12}\int \mathrm{d}^2v_1\mathrm{d}^2v_2
    \exp\left[-\frac{m\vec v_1^2}{2T_1}\right]
    \exp\left[-\frac{m\vec v_1^2}{2T_1}\right]
    g_{12}(R_{12})
    i\mathcal T^{(12)}_{12}\frac{m_1}{2}\vec v_1^2\notag
  \end{equation}
  Writing $\vec R_{12}$ in polar coordinates such that $\uvec
  R_{12}\cdot\vec v_{12} = v_{12}\cos\phi$, the radial integration is
  simply the application of the $\delta$-function in $\mathcal
  T^{(12)}_{12}$ and the step function constraints the angular
  integration.
  \begin{equation}
    \label{eq:17}
    \avr{i\Lv_{12}T_1} = \mathcal N\frac{m_1}{2}
    \int_{\pi/2}^{3\pi/2}\mathrm{d}\phi
    \int \mathrm{d}^2v_1\mathrm{d}^2v_2v_{12}\cos\phi
    \exp\left[-\frac{m_1\vec v_1^2}{2T_1(t)}\right]
    \exp\left[-\frac{m_2\vec v_2^2}{2T_2(t)}\right]
    (b_{12}^{(12)} - 1)\vec v_1^2\notag
  \end{equation}
  where
  \begin{equation}
    \label{eq:18}
    \mathcal N = x_2n(a_1 + a_2)\chi_{12}
    \left[\frac{m_1}{2\pi T_1(t)}\right]
    \left[\frac{m_2}{2\pi T_2(t)}\right]\notag
  \end{equation}
  
  According to the collision rules, the application of $b^{(12)}_{12}$
  yields
  \begin{equation}
    \label{eq:19}
    (b^{(12)}_{12} - 1)\vec v_1^2 =
    -\frac{2\mu}{m_1}(1+\epsilon_{12})(\uvec n\cdot\vec v_{12})
    (\uvec n\cdot\vec v_1)
    + \frac{\mu^2}{m_1^2}(1+\epsilon_{12})^2(\uvec n\cdot\vec v_{12})^2\notag
  \end{equation}
  where $\mu\equiv\mu_{12}$ is the reduced mass. Introducing the new
  average
  \begin{equation}
    \label{eq:23}
    \avr A_2 := \int_{\pi/2}^{3\pi/2}\mathrm{d}\phi
    \int \mathrm{d}^2v_1\mathrm{d}^2v_2v_{12}\cos\phi A
    \exp\left[-\frac{m_1\vec v_1^2}{2T_1(t)}\right]
    \exp\left[-\frac{m_2\vec v_2^2}{2T_2(t)}\right]\notag
  \end{equation}
  what we have to calculate is
  \begin{equation}
    \label{eq:24}
    \avr{i\Lv_{12}T_1} = -\mu\mathcal N(1+\epsilon_{12})
    \avr{(\uvec n\cdot\vec v_{12})(\uvec n\cdot\vec v_1)}_2
    + \frac{\mu^2}{2m_1}\mathcal N(1+\epsilon_{12})^2
    \avr{(\uvec n\cdot\vec v_{12})^2}_2
  \end{equation}
  Let's consider the first term in eq.~\eqref{eq:24}. Substituting
  $\vec v\equiv\vec v_{12}$ for $\vec v_2$ and writing $\vec v_1$ in
  polar coordinates such that $\vec v_1\cdot\vec v = v_1v\cos\gamma$
  one gets
  \begin{equation}
    \label{eq:25}
    \begin{aligned}
      \avr{(\vec v_{12}\cdot\uvec n)(\vec v_1\cdot\uvec n)}_2
      = \int &\mathrm{d}^2v\int_{\pi/2}^{3\pi/2}d\phi
      \int_0^{\infty}dv_1\int_0^{2\pi}d\gamma v^2v_1^2
      \cos^2\phi\cos(\gamma-\phi)\\
      &\times\exp\left[-\frac 12
        \frac{m_2T_1(t) + m_1T_2(t)}
        {T_1(t)T_2(t)}v_1^2
      \right]
      \exp\left[-\frac{m_2\vec v^2}{2T_2(t)}\right]
      \exp\left[\frac{m_2v_1v}{T_2(t)}\cos\gamma\right]
    \end{aligned}\notag
  \end{equation}
  Invoking the addition theorem for $\cos(\gamma-\phi)$ the
  integration over $\phi$ becomes trivial and the integration over
  $\gamma$ defines the associated Bessel function $I_1(x)$.
  \begin{equation}
    \label{eq:26}
    \avr{(\vec v_{12}\cdot\uvec n)(\vec v_1\cdot\uvec n)}_2
    = -\frac{8\pi}{3}\int \mathrm{d}^2v\int_0^{\infty}\mathrm{d}v_1v^2v_1^2
    I_1(m_2vv_1/T_2(t))
    \exp\left[-\frac 12
      \frac{m_2T_1(t) + m_1T_2(t)}
      {T_1(t)T_2(t)}v_1^2
    \right]
    \exp\left[-\frac{m_2\vec v^2}{2T_2(t)}\right]\notag
  \end{equation}
  Integrals of the form $\int \mathrm{d}xx^{n+1}I_n(\alpha
  x)\exp(-\beta x^2)$ have closed solutions such that we get
  \begin{equation}
    \label{eq:27}
    \avr{(\vec v_{12}\cdot\uvec n)(\vec v_1\cdot\uvec n)}_2
    = -\frac{8\pi}{3}\frac{m_2}{T_2(t)}
    \left[\frac{T_1(t)T_2(t)}{m_2T_1(t) + m_1T_2(t)}\right]^2
    \int \mathrm{d}^2vv^3\exp\left[
      -\frac{m_1m_2}{2}
      \frac{\vec v^2}{m_2T_1(t) + m_1T_2(t)}
    \right]\notag
  \end{equation}
  We are left with a pair of Gaussian integrals. Calculating the
  second term in eq.~\eqref{eq:24} involves essentially the same steps
  as shown above.
\end{widetext}

\section{Solving the integral equations}
\label{sec:solv-integr-equat}

To be able to apply Banach's fix point iteration, we rearrange
eqs. \eqref{eq:integral_temp} and \eqref{eq:integral_omega} and define
operators
\begin{widetext} 
  \begin{equation}
    \label{banach_T}
    A_1[T](a) :=
    \frac{-C \int \limits_{R_1}^{R_2}
      \frac{r^3}{r^3+a^3}(a+r)^2
      \sqrt{\frac{T(a)}{a^3}+\frac{T(r)}{r^3}}  (1+\epsilon)^2
      \frac{a^3}{a^3+r^3}T(r)\mathrm{d}r- H(a)}
    {
      C
      \int \limits_{R_1}^{R_2} \frac{r^3}{r^3+a^3}(a+r)^2
      \sqrt{\frac{T(a)}{a^3}+\frac{T(r)}{r^3}}\left[(\epsilon^2-1)-
        (1+\epsilon)^2\frac{a^3}{a^3+r^3}\right]\mathrm{d}r}\notag
  \end{equation}
  and
  \begin{equation}
    \label{banach_omega}
    A_2[\omega_0^{-2}](a):=
    \frac{-C
      \int \limits_{R_1}^{R_2}\frac{r^3}{r^3+a^3}(a+r)^2
      \sqrt{\frac{\omega_0^{-2}(a)}{a^3}+\frac{\omega_0^{-2}(r)}{r^3}}
      (1+\epsilon)^2
      \frac{a^3}{a^3+r^3}\omega^{-2}(r)\mathrm{d}r -2\omega_0^{-2}(a)}
    {
      C \int \limits_{R_1}^{R_2}\frac{r^3}{r^3+a^3}(a+r)^2
      \sqrt{\frac{\omega_0^{-2}(a)}{a^3}+\frac{\omega_0^{-2}(r)}{r^3}}
      \left[(\epsilon^2-1)-
        (1+\epsilon)^2\frac{a^3}{a^3+r^3}\right] \mathrm{d}r}\notag
  \end{equation}
\end{widetext}
with $C = n\sqrt6/(R_2-R_1)$. Now the solutions
of the integral equations are the fix points of $A_1$ and $A_2$, which
we try to determine by iteration.  This method worked well in the case
of eq. \eqref{eq:integral_temp}, for $\omega_0$, however, convergence
was not fully satisfactory.

That is why we combined it with Newtons method. 
We define the function $G$ whose root is to be determined by
\begin{widetext}
  \begin{equation}
    G[f](a)= 2 f(a) + C
    \int \limits_{R_1}^{R_2} \frac{r^3}{r^3 + a^3} (a+r)^2
    \sqrt{\frac{f(a)}{a^3} + \frac{f(r)}{r^3}} 
    \left[(\epsilon^2-1) f(a) + (1+\epsilon)^2
      \frac{a^3}{a^3+r^3} (f(r)-f(a))\right] \mathrm{d}r\notag
  \end{equation}
\end{widetext}
and calculate its functional derivative. After discretization of the
integrals we obtain a function $G: \mathbbm{R}^M \rightarrow
\mathbbm{R}^M$ on which we can apply Newton's method. Newton's method
requiring a sufficiently good starting approximation, we chose as such
the result of Banach's fixpoint iteration after about 300 iterations.

\bibliography{literatur}

\begin{thebibliography}{53}
\expandafter\ifx\csname natexlab\endcsname\relax\def\natexlab#1{#1}\fi
\expandafter\ifx\csname bibnamefont\endcsname\relax
  \def\bibnamefont#1{#1}\fi
\expandafter\ifx\csname bibfnamefont\endcsname\relax
  \def\bibfnamefont#1{#1}\fi
\expandafter\ifx\csname citenamefont\endcsname\relax
  \def\citenamefont#1{#1}\fi
\expandafter\ifx\csname url\endcsname\relax
  \def\url#1{\texttt{#1}}\fi
\expandafter\ifx\csname urlprefix\endcsname\relax\def\urlprefix{URL }\fi
\providecommand{\bibinfo}[2]{#2}
\providecommand{\eprint}[2][]{\url{#2}}

\bibitem[{\citenamefont{Shinbrot and Muzzio}(2000)}]{shinbrot+muzzio00}
\bibinfo{author}{\bibfnamefont{T.}~\bibnamefont{Shinbrot}} \bibnamefont{and}
  \bibinfo{author}{\bibfnamefont{F.~J.} \bibnamefont{Muzzio}},
  \bibinfo{journal}{Phys. Today} \textbf{\bibinfo{volume}{53}},
  \bibinfo{pages}{25} (\bibinfo{year}{2000}).

\bibitem[{\citenamefont{Aranson and Tsimring}(2006)}]{aranson+tsimring06}
\bibinfo{author}{\bibfnamefont{I.~S.} \bibnamefont{Aranson}} \bibnamefont{and}
  \bibinfo{author}{\bibfnamefont{L.~S.} \bibnamefont{Tsimring}},
  \bibinfo{journal}{Rev. Mod. Phys.} \textbf{\bibinfo{volume}{78}},
  \bibinfo{eid}{641} (\bibinfo{year}{2006}).

\bibitem[{\citenamefont{Goldhirsch}(2003)}]{goldhirsch03}
\bibinfo{author}{\bibfnamefont{I.}~\bibnamefont{Goldhirsch}},
  \bibinfo{journal}{Annu. Rev. Fluid Mech.} \textbf{\bibinfo{volume}{35}},
  \bibinfo{pages}{267} (\bibinfo{year}{2003}).

\bibitem[{\citenamefont{Jenkins and Mancini}(1987)}]{jenkins+mancini87}
\bibinfo{author}{\bibfnamefont{J.~T.} \bibnamefont{Jenkins}} \bibnamefont{and}
  \bibinfo{author}{\bibfnamefont{F.}~\bibnamefont{Mancini}},
  \bibinfo{journal}{J. Appl. Mech.} \textbf{\bibinfo{volume}{54}},
  \bibinfo{pages}{27} (\bibinfo{year}{1987}).

\bibitem[{\citenamefont{Jenkins and Mancini}(1989)}]{jenkins+mancini89}
\bibinfo{author}{\bibfnamefont{J.~T.} \bibnamefont{Jenkins}} \bibnamefont{and}
  \bibinfo{author}{\bibfnamefont{F.}~\bibnamefont{Mancini}},
  \bibinfo{journal}{Phys. Fluids A} \textbf{\bibinfo{volume}{1}},
  \bibinfo{pages}{2050} (\bibinfo{year}{1989}).

\bibitem[{\citenamefont{Garz{\'o} and Dufty}(1999)}]{garzo+dufty99}
\bibinfo{author}{\bibfnamefont{V.}~\bibnamefont{Garz{\'o}}} \bibnamefont{and}
  \bibinfo{author}{\bibfnamefont{J.}~\bibnamefont{Dufty}},
  \bibinfo{journal}{Phys. Rev. E} \textbf{\bibinfo{volume}{60}},
  \bibinfo{pages}{5706} (\bibinfo{year}{1999}).

\bibitem[{\citenamefont{Lu et~al.}(2000)\citenamefont{Lu, Liu, Bie, Yang, and
  Gidaspow}}]{lu+liu00}
\bibinfo{author}{\bibfnamefont{H.~L.} \bibnamefont{Lu}},
  \bibinfo{author}{\bibfnamefont{W.~T.} \bibnamefont{Liu}},
  \bibinfo{author}{\bibfnamefont{R.~S.} \bibnamefont{Bie}},
  \bibinfo{author}{\bibfnamefont{L.~D.} \bibnamefont{Yang}}, \bibnamefont{and}
  \bibinfo{author}{\bibfnamefont{D.}~\bibnamefont{Gidaspow}},
  \bibinfo{journal}{Physica A} \textbf{\bibinfo{volume}{284}},
  \bibinfo{pages}{265} (\bibinfo{year}{2000}).

\bibitem[{\citenamefont{Barrat and Trizac}(2002)}]{barrat+trizac02}
\bibinfo{author}{\bibfnamefont{A.}~\bibnamefont{Barrat}} \bibnamefont{and}
  \bibinfo{author}{\bibfnamefont{E.}~\bibnamefont{Trizac}},
  \bibinfo{journal}{Granular Matter} \textbf{\bibinfo{volume}{4}},
  \bibinfo{pages}{57} (\bibinfo{year}{2002}).

\bibitem[{\citenamefont{Dahl et~al.}(2002{\natexlab{a}})\citenamefont{Dahl,
  Hrenya, Garz{\'o}, and Dufty}}]{dahl+hrenya02}
\bibinfo{author}{\bibfnamefont{S.~R.} \bibnamefont{Dahl}},
  \bibinfo{author}{\bibfnamefont{C.~M.} \bibnamefont{Hrenya}},
  \bibinfo{author}{\bibfnamefont{V.}~\bibnamefont{Garz{\'o}}},
  \bibnamefont{and} \bibinfo{author}{\bibfnamefont{J.~W.} \bibnamefont{Dufty}},
  \bibinfo{journal}{Phys. Rev. E} \textbf{\bibinfo{volume}{66}},
  \bibinfo{pages}{041301} (\bibinfo{year}{2002}{\natexlab{a}}).

\bibitem[{\citenamefont{Pagnani et~al.}(2002)\citenamefont{Pagnani, Marconi,
  and Puglisi}}]{pagnani+marconi02}
\bibinfo{author}{\bibfnamefont{R.}~\bibnamefont{Pagnani}},
  \bibinfo{author}{\bibfnamefont{U.~M.~B.} \bibnamefont{Marconi}},
  \bibnamefont{and} \bibinfo{author}{\bibfnamefont{A.}~\bibnamefont{Puglisi}},
  \bibinfo{journal}{Phys. Rev. E} \textbf{\bibinfo{volume}{66}},
  \bibinfo{pages}{051304} (\bibinfo{year}{2002}).

\bibitem[{\citenamefont{Alam and Luding}(2003)}]{alam+luding03}
\bibinfo{author}{\bibfnamefont{M.}~\bibnamefont{Alam}} \bibnamefont{and}
  \bibinfo{author}{\bibfnamefont{S.}~\bibnamefont{Luding}},
  \bibinfo{journal}{J. Fluid. Mech.} \textbf{\bibinfo{volume}{476}},
  \bibinfo{pages}{69} (\bibinfo{year}{2003}).

\bibitem[{\citenamefont{Galvin et~al.}(2005)\citenamefont{Galvin, Dahl, and
  Hrenya}}]{galvin+dahl05}
\bibinfo{author}{\bibfnamefont{J.~E.} \bibnamefont{Galvin}},
  \bibinfo{author}{\bibfnamefont{S.~R.} \bibnamefont{Dahl}}, \bibnamefont{and}
  \bibinfo{author}{\bibfnamefont{C.~M.} \bibnamefont{Hrenya}},
  \bibinfo{journal}{J. Fluid. Mech.} \textbf{\bibinfo{volume}{528}},
  \bibinfo{pages}{207} (\bibinfo{year}{2005}).

\bibitem[{\citenamefont{Alam and Luding}(2005)}]{alam+luding05}
\bibinfo{author}{\bibfnamefont{M.}~\bibnamefont{Alam}} \bibnamefont{and}
  \bibinfo{author}{\bibfnamefont{S.}~\bibnamefont{Luding}},
  \bibinfo{journal}{Phys. Fluids} \textbf{\bibinfo{volume}{17}},
  \bibinfo{pages}{063303} (\bibinfo{year}{2005}).

\bibitem[{\citenamefont{Garz{\'o} and Montanero}(2007)}]{garzo+montanero07}
\bibinfo{author}{\bibfnamefont{V.}~\bibnamefont{Garz{\'o}}} \bibnamefont{and}
  \bibinfo{author}{\bibfnamefont{J.~M.} \bibnamefont{Montanero}},
  \bibinfo{journal}{J. Stat. Phys.} \textbf{\bibinfo{volume}{129}},
  \bibinfo{pages}{27} (\bibinfo{year}{2007}).

\bibitem[{\citenamefont{Losert et~al.}(1999)\citenamefont{Losert, Cooper,
  Delour, Kudrolli, and Gollub}}]{losert+cooper99}
\bibinfo{author}{\bibfnamefont{W.}~\bibnamefont{Losert}},
  \bibinfo{author}{\bibfnamefont{D.~G.~W.} \bibnamefont{Cooper}},
  \bibinfo{author}{\bibfnamefont{J.}~\bibnamefont{Delour}},
  \bibinfo{author}{\bibfnamefont{A.}~\bibnamefont{Kudrolli}}, \bibnamefont{and}
  \bibinfo{author}{\bibfnamefont{J.~P.} \bibnamefont{Gollub}},
  \bibinfo{journal}{Chaos} \textbf{\bibinfo{volume}{9}}, \bibinfo{pages}{682}
  (\bibinfo{year}{1999}).

\bibitem[{\citenamefont{Zamankhan}(1995)}]{zamankha95}
\bibinfo{author}{\bibfnamefont{P.}~\bibnamefont{Zamankhan}},
  \bibinfo{journal}{Phys. Rev. E} \textbf{\bibinfo{volume}{52}},
  \bibinfo{pages}{4877} (\bibinfo{year}{1995}).

\bibitem[{\citenamefont{Dahl et~al.}(2002{\natexlab{b}})\citenamefont{Dahl,
  Clelland, and Hrenya}}]{dahl+clelland02}
\bibinfo{author}{\bibfnamefont{S.~R.} \bibnamefont{Dahl}},
  \bibinfo{author}{\bibfnamefont{R.}~\bibnamefont{Clelland}}, \bibnamefont{and}
  \bibinfo{author}{\bibfnamefont{C.~M.} \bibnamefont{Hrenya}},
  \bibinfo{journal}{Phys. Fluids.} \textbf{\bibinfo{volume}{14}},
  \bibinfo{pages}{1972} (\bibinfo{year}{2002}{\natexlab{b}}).

\bibitem[{\citenamefont{Iddir and Arastoopour}(2005)}]{iddir+arastoopour05}
\bibinfo{author}{\bibfnamefont{H.}~\bibnamefont{Iddir}} \bibnamefont{and}
  \bibinfo{author}{\bibfnamefont{H.}~\bibnamefont{Arastoopour}},
  \bibinfo{journal}{AIChE J.} \textbf{\bibinfo{volume}{51}},
  \bibinfo{pages}{1620} (\bibinfo{year}{2005}).

\bibitem[{\citenamefont{Lambiotte and Brenig}(2005)}]{lambiotte+brennig05}
\bibinfo{author}{\bibfnamefont{R.}~\bibnamefont{Lambiotte}} \bibnamefont{and}
  \bibinfo{author}{\bibfnamefont{L.}~\bibnamefont{Brenig}},
  \bibinfo{journal}{Phys. Rev. E} \textbf{\bibinfo{volume}{72}},
  \bibinfo{pages}{042301} (\bibinfo{year}{2005}).

\bibitem[{\citenamefont{Garz{\'o}
  et~al.}(2007{\natexlab{a}})\citenamefont{Garz{\'o}, Dufty, and
  Hrenya}}]{garzo+dufty07}
\bibinfo{author}{\bibfnamefont{V.}~\bibnamefont{Garz{\'o}}},
  \bibinfo{author}{\bibfnamefont{J.~W.} \bibnamefont{Dufty}}, \bibnamefont{and}
  \bibinfo{author}{\bibfnamefont{C.~M.} \bibnamefont{Hrenya}},
  \bibinfo{journal}{Phys. Rev. E} \textbf{\bibinfo{volume}{76}},
  \bibinfo{pages}{031303} (\bibinfo{year}{2007}{\natexlab{a}}); a)
  \bibinfo{author}{\bibfnamefont{V.}~\bibnamefont{Garz{\'o}}},
  \bibinfo{author}{\bibfnamefont{C.~M.} \bibnamefont{Hrenya}},
  \bibnamefont{and} \bibinfo{author}{\bibfnamefont{J.~W.} \bibnamefont{Dufty}},
  \bibinfo{journal}{Phys. Rev. E} \textbf{\bibinfo{volume}{76}},
  \bibinfo{pages}{031304} (\bibinfo{year}{2007}{\natexlab{b}}).

\bibitem[{\citenamefont{Zhi-Yuan and Duan-Ming}(2008)}]{zhi-yuan+duan-ming08}
\bibinfo{author}{\bibfnamefont{C.}~\bibnamefont{Zhi-Yuan}} \bibnamefont{and}
  \bibinfo{author}{\bibfnamefont{Z.}~\bibnamefont{Duan-Ming}},
  \bibinfo{journal}{Chinese Phys. Lett.} \textbf{\bibinfo{volume}{25}},
  \bibinfo{pages}{1583} (\bibinfo{year}{2008}).

\bibitem[{\citenamefont{Williams and MacKintosh}(1996)}]{williams+mackintosh96}
\bibinfo{author}{\bibfnamefont{D.~R.~M.} \bibnamefont{Williams}}
  \bibnamefont{and} \bibinfo{author}{\bibfnamefont{F.~C.}
  \bibnamefont{MacKintosh}}, \bibinfo{journal}{Phys. Rev. E}
  \textbf{\bibinfo{volume}{54}}, \bibinfo{pages}{R9} (\bibinfo{year}{1996}).

\bibitem[{\citenamefont{Prevost et~al.}(2002)\citenamefont{Prevost, Egolf, and
  Urbach}}]{prevost+egolf02}
\bibinfo{author}{\bibfnamefont{A.}~\bibnamefont{Prevost}},
  \bibinfo{author}{\bibfnamefont{D.~A.} \bibnamefont{Egolf}}, \bibnamefont{and}
  \bibinfo{author}{\bibfnamefont{J.~S.} \bibnamefont{Urbach}},
  \bibinfo{journal}{Phys. Rev. Lett.} \textbf{\bibinfo{volume}{89}},
  \bibinfo{pages}{084301} (\bibinfo{year}{2002}).

\bibitem[{\citenamefont{Aranson and Olafsen}(2002)}]{aranson+olafsen02}
\bibinfo{author}{\bibfnamefont{I.~S.} \bibnamefont{Aranson}} \bibnamefont{and}
  \bibinfo{author}{\bibfnamefont{J.~S.} \bibnamefont{Olafsen}},
  \bibinfo{journal}{Phys. Rev. E} \textbf{\bibinfo{volume}{66}},
  \bibinfo{pages}{061302} (\bibinfo{year}{2002}).

\bibitem[{\citenamefont{Kohlstedt et~al.}(2005)\citenamefont{Kohlstedt,
  Snezhko, Sapozhnikov, Aranson, Olafsen, and Ben-Naim}}]{kohlstedt+snezhko05}
\bibinfo{author}{\bibfnamefont{K.}~\bibnamefont{Kohlstedt}},
  \bibinfo{author}{\bibfnamefont{A.}~\bibnamefont{Snezhko}},
  \bibinfo{author}{\bibfnamefont{M.~V.} \bibnamefont{Sapozhnikov}},
  \bibinfo{author}{\bibfnamefont{I.~S.} \bibnamefont{Aranson}},
  \bibinfo{author}{\bibfnamefont{J.~S.} \bibnamefont{Olafsen}},
  \bibnamefont{and} \bibinfo{author}{\bibfnamefont{E.}~\bibnamefont{Ben-Naim}},
  \bibinfo{journal}{Phys. Rev. Lett.} \textbf{\bibinfo{volume}{95}},
  \bibinfo{pages}{068001} (\bibinfo{year}{2005}).

\bibitem[{\citenamefont{Maa{\ss} et~al.}(2008)\citenamefont{Maa{\ss}, Isert,
  Maret, and Aegerter}}]{maass+isert08}
\bibinfo{author}{\bibfnamefont{C.~C.} \bibnamefont{Maa{\ss}}},
  \bibinfo{author}{\bibfnamefont{N.}~\bibnamefont{Isert}},
  \bibinfo{author}{\bibfnamefont{G.}~\bibnamefont{Maret}}, \bibnamefont{and}
  \bibinfo{author}{\bibfnamefont{C.~M.} \bibnamefont{Aegerter}},
  \bibinfo{journal}{Phys. Rev. Lett.} \textbf{\bibinfo{volume}{100}},
  \bibinfo{pages}{248001} (\bibinfo{year}{2008}).

\bibitem[{\citenamefont{Ohja et~al.}(2004)\citenamefont{Ohja, Lemieux, Dixon,
  Liu, and Durian}}]{ohja+lemieux04}
\bibinfo{author}{\bibfnamefont{R.~P.} \bibnamefont{Ohja}},
  \bibinfo{author}{\bibfnamefont{P.-A.} \bibnamefont{Lemieux}},
  \bibinfo{author}{\bibfnamefont{P.~K.} \bibnamefont{Dixon}},
  \bibinfo{author}{\bibfnamefont{A.~J.} \bibnamefont{Liu}}, \bibnamefont{and}
  \bibinfo{author}{\bibfnamefont{D.~J.} \bibnamefont{Durian}},
  \bibinfo{journal}{Nature} \textbf{\bibinfo{volume}{427}},
  \bibinfo{pages}{521} (\bibinfo{year}{2004}).

\bibitem[{\citenamefont{Abate and Durian}(2005)}]{abate+durian05}
\bibinfo{author}{\bibfnamefont{A.~R.} \bibnamefont{Abate}} \bibnamefont{and}
  \bibinfo{author}{\bibfnamefont{D.~J.} \bibnamefont{Durian}},
  \bibinfo{journal}{Phys. Rev. E} \textbf{\bibinfo{volume}{72}},
  \bibinfo{pages}{031305} (\bibinfo{year}{2005}).

\bibitem[{\citenamefont{Schr\"oter et~al.}(2005)\citenamefont{Schr\"oter,
  Goldman, and Swinney}}]{schroeter+goldman05}
\bibinfo{author}{\bibfnamefont{M.}~\bibnamefont{Schr\"oter}},
  \bibinfo{author}{\bibfnamefont{D.~I.} \bibnamefont{Goldman}},
  \bibnamefont{and} \bibinfo{author}{\bibfnamefont{H.~L.}
  \bibnamefont{Swinney}}, \bibinfo{journal}{Phys. Rev. E}
  \textbf{\bibinfo{volume}{71}}, \bibinfo{pages}{030301}
  (\bibinfo{year}{2005}).

\bibitem[{\citenamefont{Huthmann and Zippelius}(1997)}]{huthmann+zippelius97}
\bibinfo{author}{\bibfnamefont{M.}~\bibnamefont{Huthmann}} \bibnamefont{and}
  \bibinfo{author}{\bibfnamefont{A.}~\bibnamefont{Zippelius}},
  \bibinfo{journal}{Phys. Rev. E} \textbf{\bibinfo{volume}{56}},
  \bibinfo{pages}{R6275} (\bibinfo{year}{1997}).

\bibitem[{\citenamefont{Aspelmeier et~al.}(2001)\citenamefont{Aspelmeier,
  Huthmann, and Zippelius}}]{aspelmeier_huthmann_zippelius_free_cooling}
\bibinfo{author}{\bibfnamefont{T.}~\bibnamefont{Aspelmeier}},
  \bibinfo{author}{\bibfnamefont{M.}~\bibnamefont{Huthmann}}, \bibnamefont{and}
  \bibinfo{author}{\bibfnamefont{A.}~\bibnamefont{Zippelius}}, in
  \emph{\bibinfo{booktitle}{{G}ranular {G}ases}}, edited by
  \bibinfo{editor}{\bibfnamefont{T.}~\bibnamefont{P{\"o}schel}}
  \bibnamefont{and} \bibinfo{editor}{\bibfnamefont{S.}~\bibnamefont{Luding}}
  (\bibinfo{publisher}{Springer Berlin et al.}, \bibinfo{year}{2001}), pp.
  \bibinfo{pages}{31--58}.

\bibitem[{\citenamefont{van Noije and
  Ernst}(1998)}]{noije_ernst_velocity_distributions}
\bibinfo{author}{\bibfnamefont{T.~P.~C.} \bibnamefont{van Noije}}
  \bibnamefont{and} \bibinfo{author}{\bibfnamefont{M.~H.} \bibnamefont{Ernst}},
  \bibinfo{journal}{Granular Matter} \textbf{\bibinfo{volume}{1}},
  \bibinfo{pages}{57} (\bibinfo{year}{1998}).

\bibitem[{\citenamefont{Goldshtein and Shapiro}(1995)}]{goldshtein+shapiro95}
\bibinfo{author}{\bibfnamefont{A.}~\bibnamefont{Goldshtein}} \bibnamefont{and}
  \bibinfo{author}{\bibfnamefont{M.}~\bibnamefont{Shapiro}},
  \bibinfo{journal}{J. Fluid Mech.} \textbf{\bibinfo{volume}{282}},
  \bibinfo{pages}{75} (\bibinfo{year}{1995}).

\bibitem[{\citenamefont{Brilliantov and
  P{\"o}schel}(2000)}]{brilliantov+poeschel00}
\bibinfo{author}{\bibfnamefont{N.~V.} \bibnamefont{Brilliantov}}
  \bibnamefont{and}
  \bibinfo{author}{\bibfnamefont{T.}~\bibnamefont{P{\"o}schel}},
  \bibinfo{journal}{Phys. Rev. E} \textbf{\bibinfo{volume}{61}},
  \bibinfo{pages}{2809} (\bibinfo{year}{2000}).

\bibitem[{\citenamefont{Ernst and Brito}(2003)}]{ernst+brito03}
\bibinfo{author}{\bibfnamefont{M.~H.} \bibnamefont{Ernst}} \bibnamefont{and}
  \bibinfo{author}{\bibfnamefont{R.}~\bibnamefont{Brito}}, in
  \emph{\bibinfo{booktitle}{Granular Gas Dynamics}}, edited by
  \bibinfo{editor}{\bibfnamefont{T.}~\bibnamefont{P{\"o}schel}}
  \bibnamefont{and}
  \bibinfo{editor}{\bibfnamefont{N.}~\bibnamefont{Brilliantov}}
  (\bibinfo{publisher}{Springer}, \bibinfo{year}{2003}), vol.
  \bibinfo{volume}{624} of \emph{\bibinfo{series}{Lecture Notes in Physics}},
  pp. \bibinfo{pages}{3--36}.

\bibitem[{\citenamefont{P{\"o}schel et~al.}(2006)\citenamefont{P{\"o}schel,
  Brilliantov, and Formella}}]{poeschel+brilliantov06}
\bibinfo{author}{\bibfnamefont{T.}~\bibnamefont{P{\"o}schel}},
  \bibinfo{author}{\bibfnamefont{N.~V.} \bibnamefont{Brilliantov}},
  \bibnamefont{and} \bibinfo{author}{\bibfnamefont{A.}~\bibnamefont{Formella}},
  \bibinfo{journal}{Phys. Rev. E} \textbf{\bibinfo{volume}{74}},
  \bibinfo{pages}{041302} (\bibinfo{year}{2006}).

\bibitem[{\citenamefont{Hopkins and Louge}(1991)}]{hopkins+louge91}
\bibinfo{author}{\bibfnamefont{M.~A.} \bibnamefont{Hopkins}} \bibnamefont{and}
  \bibinfo{author}{\bibfnamefont{M.~Y.} \bibnamefont{Louge}},
  \bibinfo{journal}{Phys. Fluids A} \textbf{\bibinfo{volume}{3}},
  \bibinfo{pages}{47} (\bibinfo{year}{1991}).

\bibitem[{\citenamefont{McNamara}(1993)}]{mcnamara93}
\bibinfo{author}{\bibfnamefont{S.}~\bibnamefont{McNamara}},
  \bibinfo{journal}{Phys. Fluids A} \textbf{\bibinfo{volume}{5}},
  \bibinfo{pages}{3056} (\bibinfo{year}{1993}).

\bibitem[{\citenamefont{Goldhirsch and Zanetti}(1993)}]{goldhirsch+zanetti93}
\bibinfo{author}{\bibfnamefont{I.}~\bibnamefont{Goldhirsch}} \bibnamefont{and}
  \bibinfo{author}{\bibfnamefont{G.}~\bibnamefont{Zanetti}},
  \bibinfo{journal}{Phys. Rev. Lett.} \textbf{\bibinfo{volume}{70}},
  \bibinfo{pages}{1619} (\bibinfo{year}{1993}).

\bibitem[{\citenamefont{Cattuto and Marconi}(2004)}]{cattuto+marconi04}
\bibinfo{author}{\bibfnamefont{C.}~\bibnamefont{Cattuto}} \bibnamefont{and}
  \bibinfo{author}{\bibfnamefont{U.~M.~B.} \bibnamefont{Marconi}},
  \bibinfo{journal}{Phys. Rev. Lett.} \textbf{\bibinfo{volume}{92}},
  \bibinfo{pages}{174502} (\bibinfo{year}{2004}).

\bibitem[{\citenamefont{Lubachevsky}(1991)}]{lubachevsky91}
\bibinfo{author}{\bibfnamefont{B.~D.} \bibnamefont{Lubachevsky}},
  \bibinfo{journal}{J. Comp. Phys.} \textbf{\bibinfo{volume}{94}},
  \bibinfo{pages}{255} (\bibinfo{year}{1991}).

\bibitem[{\citenamefont{Fiege et~al.}(2009)\citenamefont{Fiege, Aspelmeier, and
  Zippelius}}]{fiege09}
\bibinfo{author}{\bibfnamefont{A.}~\bibnamefont{Fiege}},
  \bibinfo{author}{\bibfnamefont{T.}~\bibnamefont{Aspelmeier}},
  \bibnamefont{and}
  \bibinfo{author}{\bibfnamefont{A.}~\bibnamefont{Zippelius}},
  \bibinfo{journal}{Phys. Rev. Lett.} \textbf{\bibinfo{volume}{102}},
  \bibinfo{pages}{098001} (\bibinfo{year}{2009}).

\bibitem[{\citenamefont{Garz\'o}(2005)}]{garzo05}
\bibinfo{author}{\bibfnamefont{V.}~\bibnamefont{Garz\'o}},
  \bibinfo{journal}{Phys. Rev. E} \textbf{\bibinfo{volume}{72}},
  \bibinfo{pages}{021106} (\bibinfo{year}{2005}).

\bibitem[{\citenamefont{Bizon et~al.}(1999)\citenamefont{Bizon, Shattuck,
  Swift, and Swinney}}]{bizon+shattuck99}
\bibinfo{author}{\bibfnamefont{C.}~\bibnamefont{Bizon}},
  \bibinfo{author}{\bibfnamefont{M.~D.} \bibnamefont{Shattuck}},
  \bibinfo{author}{\bibfnamefont{J.~B.} \bibnamefont{Swift}}, \bibnamefont{and}
  \bibinfo{author}{\bibfnamefont{H.~L.} \bibnamefont{Swinney}},
  \bibinfo{journal}{Phys. Rev. E} \textbf{\bibinfo{volume}{60}},
  \bibinfo{pages}{4340} (\bibinfo{year}{1999}).

\bibitem[{\citenamefont{Haff}(1983)}]{haff83}
\bibinfo{author}{\bibfnamefont{P.~K.} \bibnamefont{Haff}}, \bibinfo{journal}{J.
  Fluid. Mech.} \textbf{\bibinfo{volume}{134}}, \bibinfo{pages}{401}
  (\bibinfo{year}{1983}).

\bibitem[{\citenamefont{Bird}(1976)}]{bird76}
\bibinfo{author}{\bibfnamefont{G.~A.} \bibnamefont{Bird}},
  \emph{\bibinfo{title}{Molecular Gas Dynamics}} (\bibinfo{publisher}{Oxford
  University Press}, \bibinfo{address}{London}, \bibinfo{year}{1976}).

\bibitem[{\citenamefont{Brey et~al.}(1996)\citenamefont{Brey, Ruiz-Montero, and
  Cubero}}]{brey+ruiz-montero96}
\bibinfo{author}{\bibfnamefont{J.~J.} \bibnamefont{Brey}},
  \bibinfo{author}{\bibfnamefont{M.~J.} \bibnamefont{Ruiz-Montero}},
  \bibnamefont{and} \bibinfo{author}{\bibfnamefont{D.}~\bibnamefont{Cubero}},
  \bibinfo{journal}{Phys. Rev. E} \textbf{\bibinfo{volume}{54}},
  \bibinfo{pages}{3664} (\bibinfo{year}{1996}).

\bibitem[{\citenamefont{Brey et~al.}(1997)\citenamefont{Brey, Dufty, and
  Santos}}]{brey+dufty97}
\bibinfo{author}{\bibfnamefont{J.~J.} \bibnamefont{Brey}},
  \bibinfo{author}{\bibfnamefont{J.~W.} \bibnamefont{Dufty}}, \bibnamefont{and}
  \bibinfo{author}{\bibfnamefont{A.}~\bibnamefont{Santos}},
  \bibinfo{journal}{J. Stat. Phys.} \textbf{\bibinfo{volume}{87}},
  \bibinfo{pages}{1051} (\bibinfo{year}{1997}).

\bibitem[{\citenamefont{Montanero and Santos}(2000)}]{montanero+santos00}
\bibinfo{author}{\bibfnamefont{J.~M.} \bibnamefont{Montanero}}
  \bibnamefont{and} \bibinfo{author}{\bibfnamefont{A.}~\bibnamefont{Santos}},
  \bibinfo{journal}{Granular Matter} \textbf{\bibinfo{volume}{2}},
  \bibinfo{pages}{53} (\bibinfo{year}{2000}).

\bibitem[{\citenamefont{Brey et~al.}(2005)\citenamefont{Brey, Ruiz-Montero,
  Maynar, and Garc\'ia~de Soria}}]{brey+ruiz-montero05}
\bibinfo{author}{\bibfnamefont{J.~J.} \bibnamefont{Brey}},
  \bibinfo{author}{\bibfnamefont{M.~J.} \bibnamefont{Ruiz-Montero}},
  \bibinfo{author}{\bibfnamefont{P.}~\bibnamefont{Maynar}}, \bibnamefont{and}
  \bibinfo{author}{\bibfnamefont{M.~I.} \bibnamefont{Garc\'ia~de Soria}},
  \bibinfo{journal}{J. Phys.: Cond. Matt.} \textbf{\bibinfo{volume}{17}},
  \bibinfo{pages}{S2489} (\bibinfo{year}{2005}).

\bibitem[{\citenamefont{Espa{\~n}ol and Warren}(1995)}]{espanol+warren95}
\bibinfo{author}{\bibfnamefont{P.}~\bibnamefont{Espa{\~n}ol}} \bibnamefont{and}
  \bibinfo{author}{\bibfnamefont{P.}~\bibnamefont{Warren}},
  \bibinfo{journal}{Europhys. Lett.} \textbf{\bibinfo{volume}{30}},
  \bibinfo{pages}{191} (\bibinfo{year}{1995}).

\bibitem[{\citenamefont{Aspelmeier et~al.}()\citenamefont{Aspelmeier, Kranz,
  and Zippelius}}]{aspelmeier09}
\bibinfo{author}{\bibfnamefont{T.}~\bibnamefont{Aspelmeier}},
  \bibinfo{author}{\bibfnamefont{W.~T.} \bibnamefont{Kranz}}, \bibnamefont{and}
  \bibinfo{author}{\bibfnamefont{A.}~\bibnamefont{Zippelius}},
  \bibinfo{note}{in preparation}.

\end{thebibliography}

\end{document}